\documentclass[aps,plb,twocolumn,showpacs,superscriptaddress,groupedaddress,nofootinbib,preprintnumbers]{revtex4-1}
\pdfoutput=1
\usepackage{amsmath}
\usepackage{amsfonts}
\usepackage{amssymb}
\usepackage{graphicx, rotating}
\usepackage{epstopdf}
\usepackage{epsfig}
\usepackage{latexsym}
\usepackage{multirow}
\usepackage{color}
\usepackage{amsmath,amssymb}
\usepackage{slashed}
\usepackage[colorlinks=true,linkcolor=red,anchorcolor=black,citecolor=blue,filecolor=cyan,menucolor=red,runcolor=filecolor,urlcolor=blue,bookmarks=true,bookmarksnumbered=true]{hyperref}
\definecolor{rossos}{cmyk}{0,1,1,0.55}
\definecolor{bluscuro}{rgb}{0.15, 0.2, .85}
\definecolor{bluchiaro}{cmyk}{1,.3,0.,0.1}

\definecolor{rossos}{cmyk}{0,1,1,0.55}
\definecolor{bluscuro}{rgb}{0.15, 0.2, .85}
\definecolor{bluchiaro}{cmyk}{1,.3,0.,0.1}

\newcommand{\bc}{\begin{center}}
\newcommand{\ec}{\end{center}}
\newcommand{\pMET}{{\bf p}\llap{/\kern1.5pt}_T}
\newcommand{\bea}{\begin{eqnarray}}
\newcommand{\eea}{\end{eqnarray}}
\newcommand{\ignore}[1]{}
\newcommand{\be}{\begin{equation}}
\newcommand{\ee}{\end{equation}}

\def\l{\label}

\def\({\left(}
\def\){\right)}
\def\<{\langle}
\def\>{\rangle}
\def\f{\frac}
\def\be{\begin{equation}}
\def\ee{\end{equation}}
\def\bry{\begin{array}}
\def\ery{\end{array}}
\def\bes{\begin{subequations}}
\def\ees{\end{subequations}}
\def\bit{\begin{itemize}}
\def\eit{\end{itemize}}
\def\ben{\begin{enumerate}}
\def\een{\end{enumerate}}
\def\dst{\displaystyle}

\newcommand{\MET}{E\llap{/\kern1.5pt}_T}
\definecolor{grey}{rgb}{0.6,0.6,0.6}
\definecolor{fuchsia}{rgb}{1,0,1}

\begin{document}

\preprint{DFPD-2015/TH/20}

\title{ATLAS diboson excess from low scale supersymmetry breaking}

%

\author{Christoffer Petersson}
\address{Department of Fundamental Physics, Chalmers University of Technology, 412 96 G\"oteborg, Sweden}
\address{Physique Th\'eorique et Math\'ematique, Universit\'e Libre de Bruxelles, C.P. 231, 1050 Brussels, Belgium}
\address{International Solvay Institutes, Brussels, Belgium}
\author{Riccardo Torre}
\address{Dipartimento di Fisica e Astronomia, Universit\`a di Padova, and INFN Sezione di Padova, Italy}
\begin{abstract}
We provide an interpretation of the recent ATLAS diboson excess in terms of a class of supersymmetric models in which the scale of supersymmetry (SUSY) breaking is in the few TeV range. The particle responsible for the excess is the scalar superpartner of the Goldstone fermion associated with SUSY breaking, the sgoldstino. This scalar couples strongly to the Standard Model vector bosons and weakly to the fermions, with all coupling strengths determined by ratios of soft SUSY breaking parameters over the SUSY breaking scale. Explaining the ATLAS excess selects particular relations and ranges for the gaugino masses, while imposing no constraints on the other superpartner masses. Moreover, this signal hypothesis predicts a rate in the $Z\gamma$ final state that is expected to be observable at the \mbox{LHC Run II} already with a few fb$^{-1}$ of integrated luminosity.    
  
\end{abstract}

\maketitle

\section{Introduction} 
\vspace{-3mm}
The ATLAS Collaboration recently published a search for resonances in the boson tagged di-jet mass distribution, featuring an excess of events around 2\,TeV \cite{Aad:2015owa}. Despite the fact that the statistical significance of the excess (up to $3.4\sigma$ locally and $2.5\sigma$ globally) is limited, the appearance of other excesses, though less significant, in similar final states and in the same mass region, such as in the CMS search in ref.~\cite{CMSCollaboration:2014df}, motivates some theoretical effort to understand the possible origin of these fluctuations. Several papers have already appeared, aimed at explaining the excess in terms of different new physics models \cite{Fukano:2015ud,Hisano:2015gna,Franzosi:2015ts,Cheung:2015vl,Dobrescu:2015va,AguilarSaavedra:2015tw,Alves:2015tf,Gao:2015ws,Thamm:2015wd,Brehmer:2015tq,Cao:2015we,Cacciapaglia:2015uf,Anonymous:2015ul,Abe:2015ud,Carmona:2015vx,Allanach:2015tr,Chiang:2015up,Cacciapaglia:2015vk,Fukano:2015vk,Sanz:2015tp,Chen:2015wa,Omura:2015vz,Chao:2015up,Anchordoqui:2015gb,Bian:2015tv,Kim:2015vl,Lane:2015wm,Faraggi:2015vi,Low:2015tn,Liew:2015to,Terazawa:2015ul,Arnan:2015vi,Niehoff:2015vw,Fichet:2015wf} and scrutinizing the ATLAS analysis \cite{Goncalves:2015ul}. 

In the context of supersymmetry (SUSY), it is not straightforward to find an explanation of this excess. For instance, with the usual particle content of the minimal SUSY extension of the Standard Model (MSSM), there is no candidate particle that could give rise to such a signal. 
However, in the case where SUSY is broken at a low scale\footnote{The current experimental lower bound on the SUSY breaking scale is at or below 1\,TeV \cite{ATLAScollaboration:2014tl} (the exact value depends on the superpartner spectrum).},  
additional degrees of freedom, related to the spontaneous breaking of SUSY, are present and can become phenomenologically relevant. In particular, the Goldstone fermion of SUSY breaking, the goldstino, and, when SUSY is linearly realized, its scalar superpartner, the sgoldstino, can couple strongly to some of the SM particles. The interaction strengths of the goldstino and sgoldstino are determined by ratios of the usual soft SUSY breaking parameters of the MSSM over the supersymmetry breaking scale. 



In this paper we provide an interpretation of the ATLAS diboson excess in terms of a class of SUSY models where the SUSY breaking scale is in the few TeV range, with a 2\,TeV sgoldstino scalar being responsible for the excess. For different discussions concerning sgoldstino physics, see, for example, refs.~
\cite{Perazzi:2000ku,
Perazzi:2000dk,
Gorbunov:2002co,
Antoniadis:2011ve,
Bertolini:2011wj,
Petersson:2011in,
Bellazzini:2012ul,
Antoniadis:2012ui,
2013PhRvD..87a3008P,
Dudas:2012ti,Dudas:2013tc}.
The sgoldstino couples mostly to the SM vector bosons, with interaction strengths determined by ratios of the gaugino masses over the SUSY breaking scale, whereas its couplings to the SM fermions are generically suppressed.    
In what follows, we study the compatibility of this signal hypothesis with the excess, identify the relevant region of the parameter space (in terms of the gaugino masses and the SUSY breaking scale) and discuss the relations to other searches in correlated channels, such as $\gamma\gamma$ and $Z\gamma$. 

The paper is organized as follows. In Section \ref{model} we provide the sgoldstino couplings to the SM and in Section \ref{excess} we extract the relevant values for the gaugino masses and SUSY breaking scale that allow us to explain the ATLAS excess.  We describe the constraints from, and implications for, other searches in Section \ref{other analyses} and conclude in Section \ref{conclusions}.
\vspace{-3mm}

\section{The sgoldstino model}\l{model} 
\vspace{-3mm}
If SUSY is realized in Nature, since the SM particles are not mass-degenerate with their superpartners, it must be in a broken phase at low energies. A general consequence of the spontaneous breaking of (global) SUSY is the existence of a Goldstone fermion, the goldstino. 
We will assume that the goldstino resides in a gauge singlet chiral superfield, with SUSY linearly realized,
\begin{equation}
\label{X}
X=x+%
\sqrt{2}\theta \widetilde{G}+%
\theta^2 F_X%
\end{equation} 
where the auxiliary field acquires a vacuum expectation value (\textsc{vev}), $\langle F_X\rangle\,{=}\,f$, that gives the dominant contribution to supersymmetry breaking.  
The focus of this paper will be on the complex scalar superpartner of the goldstino, the sgoldstino $x$ in eq.~\eqref{X}. In contrast to the goldstino, the sgoldstino is not protected by the Goldstone theorem and therefore it will generically acquire  a mass, with a value that is model-dependent. Also, in general, the masses of the CP-even and CP-odd scalars do not need to be equal. 
Here we assume them to be equal and fix them to be 2\,TeV. 

One way to take into account the interactions of the goldstino and sgoldstino is to simply promote all the usual MSSM soft terms to SUSY operators involving the goldstino superfield in eq.~\eqref{X}. For instance, the gaugino masses $m_i$, where $i\,{=}\,1,\,2$ and 3 corresponds to the bino, wino and gluino masses, respectively, are promoted to the following SUSY operators,
 \begin{eqnarray}
\label{gaugino}
\frac{m_{i}}{2} \lambda_{(i)}^\alpha \lambda_{(i)\alpha}  \to \frac{m_{i}}{2f}\int d^2\theta \, X   \, W_{(i)}^\alpha W_{(i)\alpha}\,,
\end{eqnarray}
where $W_{(i)}^\alpha$, for $i{=}1,2$ and $3$, corresponds to the gauge field-strength superfield for $U(1)_Y$, $SU(2)_L$ and  $SU(3)_c$. Note that by taking the auxiliary component of $X$ and inserting its \textsc{vev}, $\langle F_X \rangle\,{=}\,f$, one recovers the usual gaugino mass terms. The goldstino or sgoldstino interactions are obtained by taking the fermion or scalar component of $X$. 

We will from hereon focus on the interactions of the sgoldstino $x\,{=}\,(\phi+ia)/\sqrt{2}$, where $\phi$ and $a$ are the CP-even and CP-odd real scalar components. All the relevant vertices arising from eq.~\eqref{gaugino} can now be collected and included in the following sgoldstino Lagrangian \cite{Perazzi:2000ku}
\begin{equation}
\label{L}
\mathcal{L}
=\mathcal{L}_{gg}+\mathcal{L}_{\gamma\gamma}+\mathcal{L}_{Z\gamma}+\mathcal{L}_{Z Z}+\mathcal{L}_{WW}+\mathcal{L}_{GG}\,,
\end{equation}
where
\bes
\begin{align}
& \mathcal{L}_{gg}=\dst \hspace{-1pt} \f{m_{3}}{2\sqrt{2}f}\hspace{-2pt} \left( -\phi G^{a\,\mu\nu}G^{a}_{\mu\nu}
\hspace{-1pt}+\hspace{-1pt}  a G^{a\,\mu\nu}\widetilde{G}^{a}_{\mu\nu}\right) \hspace{-1pt},
\\
& \mathcal{L}_{ WW}=\dst \hspace{-1pt}\f{m_2}{\sqrt{2}f}\hspace{-2pt}\left( -\phi W^{+\,\mu\nu}W^{-}_{\mu\nu} \hspace{-1pt}+\hspace{-1pt}
a W^{+\,\mu\nu}\widetilde{W}^{-}_{\mu\nu} \right)\hspace{-1pt},
\\
& \mathcal{L}_{  Z Z} = \hspace{-1pt}\f{m_{1} s^{2}_{\theta_W} {+}m_{2} c^{2}_{\theta_W}}{2\sqrt{2}f}\hspace{-1pt} \left(-\phi Z^{\mu\nu}Z_{\mu\nu} \hspace{-1pt}+\hspace{-1pt}
a Z^{\mu\nu}\widetilde{Z}_{\mu\nu} \right)\hspace{-1pt},
\\
& \mathcal{L}_{ \gamma\gamma}=\dst \hspace{-1pt}\f{m_{1} c^{2}_{\theta_W} {+}m_{2} s^{2}_{\theta_W} }{2\sqrt{2}f} \hspace{-1pt}\left( -\phi F^{\mu\nu}F_{\mu\nu}
\hspace{-1pt}+\hspace{-1pt} a F^{\mu\nu}\widetilde{F}_{\mu\nu} \right)\hspace{-1pt},
\\
& \mathcal{L}_{Z\gamma} = \hspace{-1pt}\f{(m_{2}{-}m_{1}) s_{\theta_W} c_{\theta_W}}{\sqrt{2}f} 
\hspace{-2pt}\left(\hspace{-1pt}-\phi F^{\mu\nu}Z_{\mu\nu} 
\hspace{-2pt}+\hspace{-1pt}  a F^{\mu\nu}\widetilde{Z}_{\mu\nu} \hspace{-1pt}\right)\hspace{-1pt},
\\
& \mathcal{L}_{ GG} = \f{m_{\phi}^2}{2\sqrt{2} f} \hspace{-1pt}\left( -\phi \,\widetilde{G}\,\widetilde{G} + i\,a\, \widetilde{G}\,\widetilde{G}\right)+ \mathrm{h.c.}\,, \label{xGG}
\end{align}
\ees
where $s_{\theta_W}{=}\sin\theta_W$ and $c_{\theta_W}{=}\cos\theta_W$, with $\theta_W$ being the weak mixing angle, and the tilde denotes e.g.~$\widetilde{G}^{a}_{\mu\nu}\,{=}\,(1/2)\epsilon_{\mu\nu\rho\sigma}G^{a\,\rho\sigma}$. The interactions in eq.\,\eqref{xGG} arise from the operator $m_{\phi}^2/(4f^2)(X^\dagger X)^2$ in the Kahler potential, from which also the soft mass $m_\phi\,{=}\,m_a$ for the CP-even and CP-odd sgoldstino scalars $\phi$ and $a$ arises.
Notice that the sgoldstino couples purely to the transverse components of the $W$ and $Z$ bosons. A small coupling to the longitudinal components can arise through mixing with the Higgs, but for the region of parameter space that we consider 
such a mixing is negligible.


From the sgoldstino Lagrangian \eqref{L} we can compute the partial decay widths for the sgoldstino scalar $\phi$ (the corresponding widths for $a$ are obtained by simply replacing $\phi\,{\to}\,a$ since they are identical to those of $\phi$),  
\bes
\begin{align}
& \bry{lll} \dst \Gamma(\phi \to gg) = \f{m_{3}^2 m^{3}_{\phi}}{4\pi f^{2}}\,,\l{widthphigg} \ery  \\
& \bry{lll} \Gamma(\phi \to WW) =  \dst \f{m_{2}^{2} m_{\phi}^3}{16\pi f^{2}} k\Big(  \frac{m_{W}}{m_\phi}  \Big)\,,
\ery\\
& \bry{lll} \Gamma(\phi \to Z Z) = \dst \f{(m_{1} s^{2}_{\theta_W} +m_{2} c^{2}_{\theta_W})^2 m^{3}_{\phi}}{32\pi f^{2}} k \Big( \frac{m_{Z}}{m_\phi} \Big)\,,
\ery 
\\
& \bry{lll} \dst  \Gamma(\phi \to \gamma\gamma) = \f{ (m_{1} c^{2}_{\theta_W} +m_{2} s^{2}_{\theta_W})^{2} m^{3}_{\phi}}{32 \pi f^{2}}\,,\l{widthphigaga} \ery\\
& \bry{lll} \dst  \Gamma(\phi \to Z\gamma) \hspace{-2pt}=\hspace{-2pt} \f{ (m_{2}{-}m_{1})^2 s_{\theta_W}^2 c_{\theta_W}^{2} m^{3}_{\phi}}{16\pi f^{2}} \bigg(1-\f{m_{Z}^{2}}{m_\phi^{2}} \bigg)^{3}, \l{widthphigaZ}
\ery\\
& \bry{lll} \dst \Gamma(\phi \to GG) = \f{m_{\phi}^5}{32\pi f^{2}}\,,\ery
\end{align}
\ees
where the function $k(x)=(1{-}4x^2{+}6x^4)(1-4x^2)^{1/2}$ is close to unity in the case where $ m_W{,}\,m_Z \ll m_\phi=2$\,TeV. 

The interactions between the sgoldstino and the SM fermions arise from superpotential operators such as $(A_u/f)  X Q H_u U^c$, which, upon taking the auxiliary component of $X$ and inserting its \textsc{vev}, also give rise to the usual $A$-terms. Since we are requiring all soft parameters to be smaller than $\sqrt{f}$, the sgoldstino couplings will be suppressed at least by the ratio of the Higgs \textsc{vev} over $\sqrt{f}$, which makes the sgoldstino decays to SM fermions negligible with respect to the sgoldstino decays to vector bosons. 


The sgoldstino is produced at the LHC by gluon-gluon fusion with the leading order production cross section (summing the two equal contributions from the CP-even scalar $\phi$ and the CP-odd scalar $a$, with $m_\phi\,{=}\,m_a$)
\be\l{prodCS}
\dst \sigma\hspace{-1pt}=\hspace{-2pt} \f{\pi^{2}\Gamma\(\phi\to gg\)}{4s m_{\phi}}\hspace{-1pt}\times\hspace{-2pt}\int_{\f{m_{\phi}^{2}}{s}}^{1}\hspace{-2pt}\f{dx}{x}f_{p/g}\hspace{-2pt}\(x,m_{\phi}^{2}\)f_{p/g}\hspace{-2pt}\(\hspace{-2pt}\f{m_{\phi}^{2}}{xs},m_{\phi}^{2}\hspace{-2pt}\)\,,
\ee
where the partial width $\Gamma\(\phi\to gg\)$ is given by Eq.~\eqref{widthphigg}, $s$ is the center of mass energy squared and $f_{p/g}\(x,Q^{2}\)$ are the parton distribution functions defined at the scale $Q^{2}$. Since $\Gamma_{\phi}/m_{\phi}$ is below $10\%$ for the sgoldstino in the relevant region of the parameter space, eq.~\eqref{prodCS}, which assumes the narrow width approximation, is always reliable. 
\vspace{-3mm}

\section{Explaining the diboson excess}\l{excess}
\vspace{-3mm}
In this section, to assess the compatibility of a sgoldstino signal with the ATLAS diboson excess, we compare the number of signal events the sgoldstino gives rise to with the number of excess events reported by ATLAS. 
Figure 5 of ref.~\cite{Aad:2015owa} shows the invariant mass distribution of the boson tagged jets for the $WZ$, $WW$ and $ZZ$ selection regions (SRs). These regions have large overlaps due to the overlap of the shapes of the single $W$ and $Z$ tagged jet mass distributions. We take into account this overlap by computing the different efficiencies $\epsilon_{V_{f}V'_{f}\to V_{s}V'_{s}}$ for a final state $V_{f}V'_{f}$ to end up in the $V_{s}V_{s}'$ SR. The values of these efficiencies are given in Table \ref{Table:Efficiencies}.

\begin{table}[t]
\begin{center}
{
\begin{tabular}{l|c|c|c}
\hline
\hline
$\qquad$  selection region & $WW$ & $WZ$ & $ZZ$\\ 
final state &&& \\ 
\hline
$WW$ & $0.39$ & $0.37$ & $0.16$ \\
$WZ$ & $0.33$ & $0.44$ & $0.25$ \\
$ZZ$ & $0.27$ & $0.47$ & $0.37$ \\
\hline
\hline
\end{tabular}
}
\\[0.1cm]
\caption{\small Efficiencies $\epsilon_{V_{f}V'_{f}\to V_{s}V'_{s}}$ for a final state $V_{f}V'_{f}$ to end up in the $V_{s}V_{s}'$ selection region.}\label{Table:Efficiencies}\vspace{-0.5cm}
\end{center}
\end{table}
We consider the window $1.75\,{-}\,2.25$ TeV 
in the boson tagged dijet mass distribution, around the mass hypothesis, and 
compare the number of observed events with the number of events predicted by the SM. A reliable combination of the three SRs $WZ$, $WW$ and $ZZ$ would require detailed knowledge of the degree of correlation between these three channels that go beyond the effect of the efficiencies $\epsilon_{}$ that we take into account, such as the correlation of all the systematic uncertainties. Since we do not have this information at our disposal, we instead extract the signal from a single channel and then confront it with the number of events observed in the other two channels, as well as with the other relevant analyses. 

Our model predicts the largest production rate in the $WW$ channel and therefore, to also minimize the uncertainties coming from our extraction of the tagging efficiencies, we extract the sgoldstino signal yield from the $WW$ SR. For an invariant mass of $2$ TeV, we obtain from the ATLAS analysis $S_{WW}=4.2_{-2.0}^{+3.2}$ 
excess events in the considered window, where the error band represents Poissonian central $68\%$ CL interval, which is what we refer to as $1\sigma$ interval throughout the paper. The number of signal events produced by the sgoldstino in the various $V_{s}V'_{s}$ SRs is given by
\be\l{eq1}
\bry{lll}
\dst S_{V_{s}V'_{s}}&\hspace{-5pt}=&\dst \hspace{-5pt}\Big[\sigma\hspace{-2pt}\times\hspace{-1pt} \text{BR}_{WW}\hspace{-2pt}\times\hspace{-2pt}\mathcal{A}_{WW}\hspace{-1pt}\times\hspace{-1pt} \text{BR}_{WW\to\text{had}}\hspace{-1pt}\times\hspace{-1pt} \epsilon_{WW\to V_{s}V'_{s}}\\
 &\hspace{-5pt}
 &\hspace{-3pt}+ \hspace{2pt}
 \dst\sigma
 \hspace{-2pt}
 \times\hspace{-1pt} \text{BR}_{ZZ}\hspace{-2pt}\times\hspace{-2pt}\mathcal{A}_{ZZ}\hspace{-1pt}\times \hspace{-1pt}\text{BR}_{ZZ\to\text{had}}\hspace{-1pt}\times \hspace{-1pt}\epsilon_{ZZ\to V_{s}V'_{s}}\Big]\mathcal{L}\,,
\ery
\ee
where $\text{BR}_{ij}$ corresponds to the sgoldstino decay branching ratio into the $ij$ final state, the factors $\mathcal{A}_{WW}$ and $\mathcal{A}_{ZZ}$ are the acceptances for the kinematic and topology selections and include the signal acceptance to the invariant mass cut in the window we consider, $\text{BR}_{WW\to\text{had}}$ and $\text{BR}_{ZZ\to\text{had}}$ are the hadronic branching ratios of the $WW$ and $ZZ$ channels, respectively, and $\mathcal{L}=20.3$ fb$^{-1}$ is the integrated luminosity.

The knowledge of the acceptance factors is a key ingredient to estimate the number of events starting from a certain theoretical value of $\sigma\times\text{BR}$. Unfortunately, the ATLAS analysis only reports the value of these acceptances for a vector, and a spin-two signal hypothesis, the bulk graviton, that decays into longitudinally polarized vector bosons. This information does not allow us to extract the acceptance for a scalar particle decaying to transverse vector bosons, as is the case of the sgoldstino. Comparing the ATLAS acceptances with the ones that CMS reports in ref.~\cite{CMSCollaboration:2014df}, which is the counterpart of ref.~\cite{Aad:2015owa}, where also a spin-two particle decaying to transverse vector bosons is considered, the RS graviton, we expect the acceptances for a resonance decaying into transverse gauge bosons to be about 50\% 
smaller than the acceptances for a resonance decaying into longitudinal vectors, which is the case reported by ATLAS. Such a reduction in the acceptances would require a larger $\sigma\times\text{BR}$ to explain the ATLAS excess, thereby selecting a region of the parameter space with lower values of the relevant parameters, namely $m_{2}$, $m_{3}$ and $\sqrt{f}$. However, since we cannot reliably estimate this number we will instead use the acceptances that ATLAS reports for the bulk graviton also for the sgoldstino. This also allows us,  in a more consistent way, to compare with other analyses where the same spin-two signal hypothesis is considered. We stress that, in the case in which a higher significance of this excess is observed in Run II, 
it is of primary importance that the experimental collaborations provide the acceptances  for all the relevant spin hypotheses and polarizations of the final state vector bosons.\\
\indent  The acceptance factors in eq.~\eqref{eq1} are estimated starting from the total selection efficiencies reported in Figure 2\,(b) of ref.~\cite{Aad:2015owa}, divided by the aforementioned boson tagging efficiencies in the respective SRs. This number is then multiplied by the efficiency corresponding to the invariant mass cut in the window we consider, estimated from the signal shape reported by ATLAS in the $WW$ SR for the spin-two resonance in Figure 5 (b), which is the one with more available statistics and which is about $0.87$. The resulting two acceptances in eq.~\eqref{eq1} are, as expected, almost identical, $\mathcal{A}_{WW}=0.22$ and $\mathcal{A}_{ZZ}=0.21$.\\
\indent The parameters that most strongly affect the diboson channels relevant for the excess are $m_{2}$, $m_{3}$ and $\sqrt{f}$, while in the case of the $\gamma\gamma$ and $Z\gamma$ channels, there is also some dependence on $m_{1}$. 
The ATLAS search in the $\gamma\gamma$ channel of ref.~\cite{ATLAScollaboration:2015uw} place a $95\%$ CL upper limit on $\sigma\times \text{BR}_{\gamma\gamma}$ at around $0.3$ fb for a mass of $2$ TeV. We stress that also this result is obtained assuming a spin-two resonance and can not be straightforwardly used to constrain a scalar. However, since we do not expect huge changes in the efficiencies, this gives us an estimate of the current bound on a heavy scalar decaying to $\gamma\gamma$. As can be seen from eq.~\eqref{widthphigaga}, this bound can be completely evaded by choosing $m_1\approx-m_2 \tan^{2}\theta_{W}$ since, in this case, $\text{BR}_{\gamma\gamma}\approx 0$. However, since we find that there is a wide range of $m_1$ that satisfies the $\gamma\gamma$ constraint, without affecting the diboson channels, we choose not to fix any particular relation to $m_2$, but instead we set it to a reference value, $m_{1}\,{=}\,100$ GeV\footnote{Notice that when $m_{1}<m_{\phi}/2$ the sgoldstino can decay into two neutralinos with a coupling that is generally model dependent. This would amount in a shift to the total width that would slightly shift the region of the parameter space where the model reproduces the ATLAS excess. This decay could also give rise to a final state with two photons and missing transverse energy that is potentially interesting at the LHC.}, for which the constraint from the $\gamma\gamma$ search is satisfied in the entire range or $m_{2}$, $m_{3}$ and $\sqrt{f}$ that we consider.


\begin{figure}[t!]
\begin{center}
\hspace{-4mm} \includegraphics[scale=0.4]{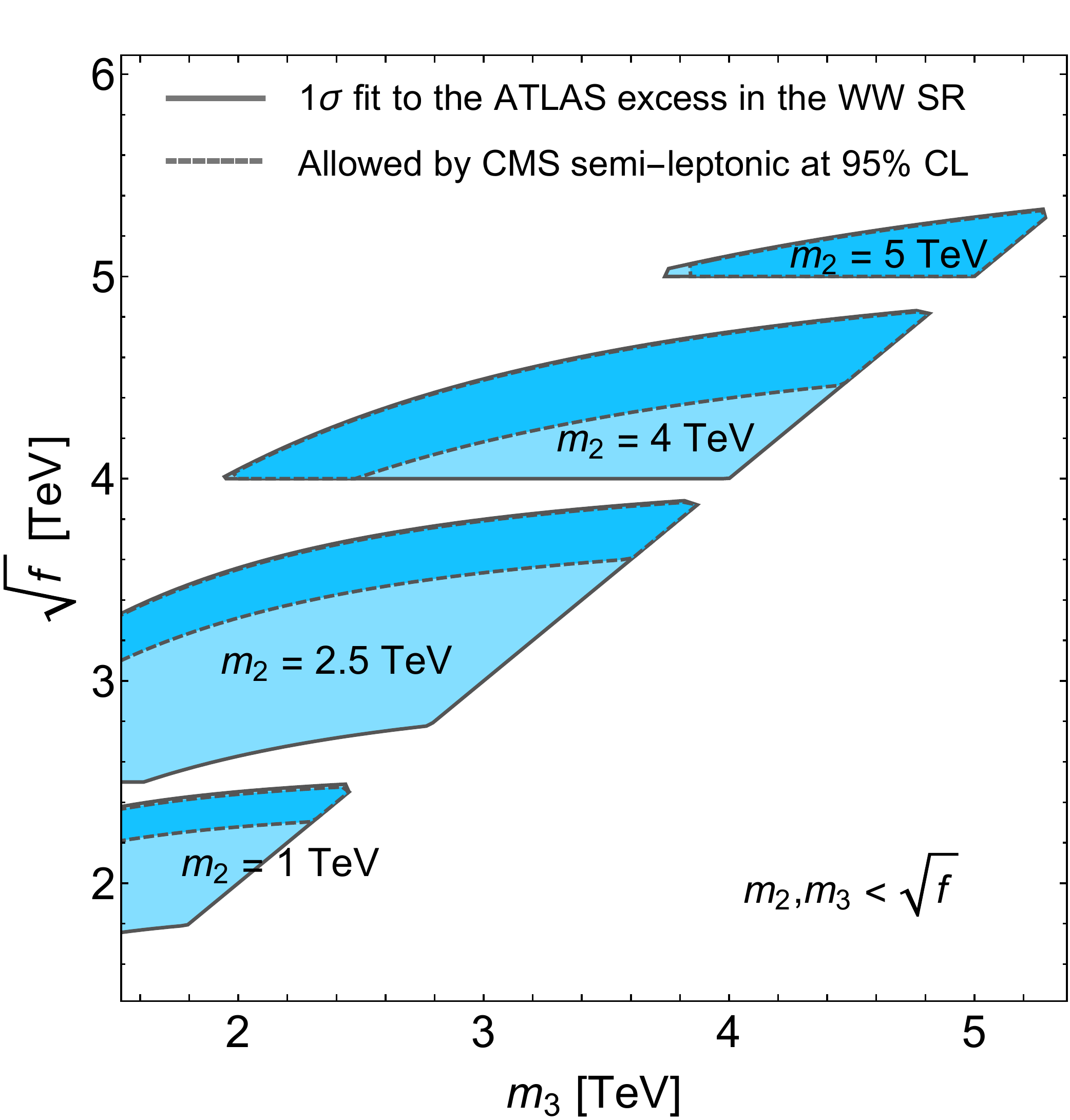}\hspace{8mm}
\vspace{-0.5cm}\caption{The four regions enclosed by the solid lines, corresponding to four different values of $m_2$, show the points in the plane $(m_{3},\sqrt{f})$ for which a 2\,TeV sgoldstino can explain the ATLAS diboson excess \cite{Aad:2015owa} within $1\sigma$. The (dark blue) regions enclosed by the dashed lines correspond to the subset of these point that satisfies the 95\% CL limit placed by the semi-leptonic diboson search by CMS \cite{CMSCollaboration:2014ke}.}\vspace{-4mm}
\label{fig1}
\end{center}
\end{figure}

The regions enclosed by the solid grey curves in Figure \ref{fig1} represent the regions in the $(m_{3},\sqrt{f})$ plane where the $2$\,TeV sgoldstino signal reproduces the number of excess events observed by ATLAS in the $WW$ SR in the invariant mass window we consider, within the $1\sigma$ band, namely $S_{WW}=4.2_{-2.0}^{+3.2}$. The four different regions correspond to four representative values of $m_2$, namely $1$, $2.5$, $4$ and $5$ TeV. Of course, different values of $m_{2}$ interpolate between these regions covering a large part of the $m_{3}\,{<}\,\sqrt{f}$ plane. In Figure \ref{fig1} we have also imposed the constraint $m_{2},m_{3}\,{<}\,\sqrt{f}$ to ensure a valid expansion in eq.~\eqref{L} in terms of effective operators. The current experimental limits on  $m_3$ and $\sqrt{f}$ depend on the masses of the other superpartners on which the fit to the ATLAS excess imposes no constraints. In Figure \ref{fig1}, we take a conservative approach and require both $m_3$ and $\sqrt{f}$ to be above 1.5\,TeV, corresponding to the most stringent current bound on both the gluino mass \cite{ATLAScollaboration:2015ue} and the SUSY breaking scale \cite{ATLAScollaboration:2014tl}.

In the four regions shown in Figure \ref{fig1}, the predicted values for the cross sections in the other two diboson channels span the ranges $\sigma\times \text{BR}_{WZ}\in [2.3,7.6]$ fb and the $\sigma\times \text{BR}_{ZZ}\in [0.7,2.2]$ fb. For these cross section intervals, by using eq.~\eqref{eq1}, we compute the $1\sigma$ intervals for the number of excess events predicted in the $WZ$ and $ZZ$ SRs, and we obtain $S_{WZ}\in [2.4,8.0]$ and $S_{ZZ}\in [1.3,4.2]$. Comparing these numbers with the corresponding values extracted from the ATLAS analysis, namely $7.0_{-2.6}^{+3.8}$ and $6.4_{-2.4}^{+3.6}$, respectively, we see that the $WZ$ SR is well within the statistical $1\sigma$ band, while $ZZ$ shows a slight tension. However, this tension is removed once one includes systematic uncertainties, which are at the level of $50\%$ for the signal \cite{Aad:2015owa}. Nevertheless, it is worth asking if we can directly understand this tension from the ATLAS analysis. The $ZZ$ SR in the ATLAS analysis is the one suffering from less statistics and is therefore expected to have the largest uncertainty. One important feature that one notices looking carefully at the ATLAS analysis is that the shape of the fitted background in the $WZ$ and $WW$ channels agree rather well, while the one in the $ZZ$ channel has a different shape at high masses, as it falls much faster. Since we do not expect substantially different shapes for these SM diboson backgrounds at such high invariant masses, this provides an estimate for the error of the fit to the background. 
In order to assess the possible origin of the tension that we find between the signal in the $ZZ$ SR and in the other two SRs, we compute the number of excess events in the $ZZ$ SR that would be obtained if we instead use the $WZ$ or the $WW$ shape, with the $ZZ$ normalization, as $ZZ$ background distribution. With the $WZ$ or the $WW$ background distribution shapes, we obtain only $3.0$ and $1.7$ excess events, respectively, in the window we consider, to be compared with the $6.4$ obtained from the ATLAS fit to the background distribution in the $ZZ$ SR. This shows that the tension in the $ZZ$ channel could be due to an inaccurate prediction of the shape of the background in the $ZZ$ SR, caused by the lack of statistics. 
\vspace{-3mm}


\section{Other analyses}\l{other analyses}
\vspace{-3mm}
Now that we have extracted the interesting region of the parameter space of the sgoldstino that allows us to reproduce the excess of events observed by ATLAS, we confront our signal hypothesis with the other relevant searches. 
The first search to compare with is reported in ref.~\cite{CMSCollaboration:2014df} and is the CMS analogous of the ATLAS fully hadronic search \cite{Aad:2015owa}. This sets the limits (for the same spin and polarization hypothesis that is considered by ATLAS and that we used to extract the signal) \mbox{$\sigma\times\text{BR}_{WW}\,{<}\,11$\,fb} and $\sigma\times\text{BR}_{ZZ}\,{<}\,10$ fb for a $2$\,TeV mass hypothesis, which lie above our $1\sigma$ bands for the corresponding quantities and hence do not set any further constraint on the allowed parameter space. 

The analysis that sets the strongest constraint on the $\sigma\times\text{BR}$ for a $2$\,TeV resonance decaying to gauge bosons is a CMS search in the semi-leptonic channel with either 1 lepton ($WW$ channel) or 2 leptons ($ZZ$ channel) \cite{CMSCollaboration:2014ke}. There is no overlap in this case due to the selection with different numbers of leptons. We can therefore directly compare our predicted $\sigma\times\text{BR}$ with the limits this search places in the respective channels, \mbox{$\sigma\times\text{BR}_{WW}\,{<}\,3$\,fb} and $\sigma\times\text{BR}_{ZZ}\,{<}\,8 $ fb at $2$ TeV. While the $ZZ$ bound does not constrain our parameter space, we get a constraint from the $WW$ channel, which reduces the allowed parameter space in the $(m_{3},\sqrt{f})$ plane in Figure \ref{fig1} to the (dark blue) regions enclosed by the dashed lines. The (light blue) regions that remain outside the dashed contours are excluded at $95\%$ CL by the CMS semi-leptonic analysis \cite{CMSCollaboration:2014ke} in the $WW$ channel.\\
\indent Let us finally comment on other possible interesting channels. If the ATLAS diboson excess is caused by the scalar sgoldstino, no signal is expected in the $ZH$ and $WH$ channels.
Hence, if statistically significant excesses are found in Run II in these channels, it 
would point toward other new physics scenarios.  
Instead, the most relevant other channels for the sgoldstino signal hypothesis are the $\gamma\gamma$ and $Z\gamma$ channels. 
As was discussed in the previous section, the bound from existing $\gamma\gamma$ searches can always be satisfied by choosing $m_1$ to be within a rather wide range around the value $m_1\approx-m_2 \tan^{2}\theta_{W}$, for which  $\text{BR}_{\gamma\gamma}$ vanishes, as can be seen from eq.~\eqref{widthphigaga}. Clearly, the allowed range of $m_1$ is wider for larger values of $m_2$. \\
\indent One way to place a constraint on $m_1$ would be to search for a resonance in the $Z\gamma$ channel at 2\,TeV, and use the relation between $m_1$ and $m_2$ in eq.~\eqref{widthphigaZ}. The only search in the $Z\gamma$ channel that we are aware of is the ATLAS analysis in ref.~\cite{ATLAScollaboration:2014ur} which only extends to invariant masses up to $1.6$ TeV. Therefore we do not get any constraint from this search. However, it is interesting to note that the exclusion at $1.6$ TeV is $\sigma\times\text{BR}_{Z\gamma}<0.17$\,fb for a scalar. We find that, in most of the the parameter space that explains the ATLAS diboson excess, and for a large range of values of $m_1$ within the region allowed by $\gamma\gamma$ searches, the $\sigma\times\text{BR}_{Z\gamma}$ we get for the 2 TeV sgoldstino is larger than 0.17\,fb. This suggests that once the $Z\gamma$ search is extended to include $2\,$TeV invariant masses, this channel will be sensitive to the sgoldstino signal and could quickly lead to a discovery or exclude most of the parameter space presently allowed.
\vspace{-3mm}
%
%

\section{Conclusions}\l{conclusions}
\vspace{-3mm}
In this paper we have provided an explanation of the recently reported ATLAS diboson excess in terms of 
a 2\,TeV sgoldstino scalar, which is present in a class of supersymmetric models in which the supersymmetry breaking scale is in the few TeV range. Fitting this excess selects particular ranges and relations among the gaugino masses, while imposing no constraints on the other superpartner masses. In terms of other resonance searches, while no signal is expected in the $ZH$ and $WH$ channels, we expect the most sensitive channel to be $Z\gamma$. \\
\indent The sgoldstino production cross-section, which originates from gluon-gluon fusion, is expected to increase by a factor of about 19 when going from $\sqrt{s}=8\,$TeV to 13\,TeV for a mass of $2$ TeV. This should be contrasted with, for example, the factor of about 7 increase of the production cross-section that is expected for a $q\bar{q}$ resonance of the same mass. Hence, with the sgoldstino signal hypothesis, also taking into account that the background is mainly due to $q\bar{q}$, one expects the diboson excess to grow significantly faster with the incoming 13\,TeV data, with respect to, for instance, a heavy vector signal hypothesis. The different scaling of the signal cross sections with the collider energy could help, in case of discovery, to understand the nature of the new resonance.\\
\indent Let us end by encouraging the ATLAS and CMS collaborations to provide the efficiencies for all the relevant spin hypotheses of the resonance and the polarizations of the vector bosons in the final state. The fact that the ATLAS analysis \cite{Aad:2015owa} only provides the efficiencies for spin-one and -two resonances decaying to longitudinally polarized vector bosons introduces a large uncertainty in our interpretation of the excess. However, we expect this to only amount to a rescaling and possibly a shift of the relevant parameter space region towards slightly lower values of $m_{2}$, $m_{3}$ and $\sqrt{f}$. \vspace{3mm}\\
\noindent {\bf Acknowledgments:}
We are thankful to A.\,Wulzer and F.\,Zwirner for discussions and comments on the manuscript. The work of C.\,P.~is supported by the Swedish Research Council (VR) under the contract 637-2013-475, by IISN-Belgium (conventions 4.4511.06, 4.4505.86 and 4.4514.08) and by the ``Communaut\'e Fran\c{c}aise de Belgique" through the ARC program and by a ``Mandat d'Impulsion Scientifique" of the F.R.S.-FNRS. 
The work of R.T. was supported by the Italian PRIN no.~2010YJ2NYW$\_$003. R.T. acknowledges CERN hospitality during the completion of this work.


\bibliographystyle{mine}
\bibliography{bibliography}

\end{document}